\documentclass[journal,twocolumn]{IEEEtran}
\usepackage{epsfig,makeidx,color}
\usepackage{amsmath,amssymb,bbm}
\usepackage{cite,graphicx,lipsum}
\usepackage{enumerate}
\usepackage[switch,pagewise]{lineno}
\usepackage{hyperref}
\hypersetup{
        colorlinks = true,
        citecolor=blue,
}
\def\rT{{\rm T}}

\def\uZ{{\mathbb Z}}

\def\uE{{\mathbb E}}
\def\uW{{\mathbb W}}

\newtheorem{mytheorem}{\bf Theorem} 
\newtheorem{mylemma}{\bf Lemma} 

\def\deft{ \buildrel \triangle \over = }

\def\eeq{ \buildrel \cdot \over = }
\def\be{ \begin{equation} }
\def\ee{ \end{equation} }
\def\bea{ \begin{eqnarray} }
\def\eea{ \end{eqnarray} }

\def\bx{{\bf x}}

\def\bq{{\bf q}}

\def\b0{{\bf 0}}

\ifCLASSOPTIONonecolumn
  \newcommand{\figwidth}{0.50\columnwidth}
  
\else
  \newcommand{\figwidth}{0.80\columnwidth}
  
\fi

\begin{document}

\title{On Fast Retrial for Two-Step Random Access in MTC}

\author{Jinho Choi\\
\thanks{The author is with
the School of Information Technology,
Deakin University, Geelong, VIC 3220, Australia
(e-mail: jinho.choi@deakin.edu.au).
This research was supported
by the Australian Government through the Australian Research
Council's Discovery Projects funding scheme (DP200100391).}}


\maketitle
\begin{abstract}
In machine-type communication (MTC), a group of devices or
sensors may need to send their data packets
with certain access delay limits for delay-sensitive
applications or real-time Internet-of-Things (IoT) applications.
In this case, 2-step random access approaches would be 
preferable to 4-step random access approaches that are employed
for most MTC standards in cellular systems.
While 2-step approaches are efficient in terms of access delay,
their access delay is still dependent on re-transmission strategies.
Thus, for a low access delay, fast retrial 
that allows immediate re-transmissions
can be employed as
a re-transmission strategy.
In this paper, we study 2-step 
random access approaches with fast retrial
as a buffered multichannel ALOHA with fast retrial,
and derive an analytical way to obtain
the quality-of-service (QoS) exponent 
for the distribution of queue length so that 
key parameters can be decided to meet QoS requirements
in terms of access delay. 
Simulation results confirm that the derived 
analytical approach can provide
a good approximation of QoS exponent.
\end{abstract}


\begin{IEEEkeywords}
Grant-free Random Access; Fast Retrial; Low Access Delay
\end{IEEEkeywords}

\ifCLASSOPTIONonecolumn
\baselineskip 30pt
\fi

\section{Introduction}	\label{S:Intro}


There has been a surge of interest in
the Internet-of-Things (IoT) 
\cite{Gubbi13} \cite{Fuqaha15} \cite{Buyya16}
in various aspects including IoT connectivity 
\cite{Qadir18} \cite{Famaey18} \cite{Ding_20Access}.
While there exist a large number of devices
to be connected, there are some devices
that require low access delay for 
mission critical applications.
To support a large number of devices,
machine-type communication (MTC)
has been studied in cellular systems
(e.g., 5th generation (5G) systems)
\cite{Bockelmann16}.
For low access delay applications,
ultra-reliable low latency communication (URLLC)
is also extensively investigated
\cite{Bennis18} \cite{Popovski19}.

In MTC, random access approaches
are employed to support a number of devices with sporadic traffic 
due to low signaling overhead. In particular, 
handshaking processes based on random access
are considered for MTC 
in standards \cite{3GPP_MTC} \cite{3GPP_NBioT}.
In general, a handshaking process
consists of 4 steps and it is often called 4-step approach.
In the first step, a device with data to send is to 
randomly choose a preamble from a pool of pre-determined preambles,
which is shared by all the devices in a cell,
and transmit it so that a base station (BS)
can allocate a dedicated uplink channel 
to the device.

There are also other approaches. For example, 2-step 
random access approaches are extensively studied
\cite{Bockelmann18}, which are also often called 
compressive random access or one-shot
\cite{Wunder14} \cite{Wunder15} \cite{Choi17IoT} \cite{Choi20b}
or grant-free approaches \cite{Abebe17} 
\cite{Senel18} \cite{Liu18} \cite{Ding20b}.
There are several differences between
4-step and 2-step approaches.
In 4-step approaches, two different uplink channels
are used, namely physical random access channel (PRACH)
and physical uplink shared channel (PUSCH).
Preambles are transmitted through PRACH, while
data packets are transmitted through PUSCH.
On the other hand, in 2-step approaches,
since both preamble and data packets are transmitted
together, one physical channel can be used.
In addition, in 2-step approaches,
the length of data blocks is likely fixed \cite{Choi20b}.
Thus, 2-step approaches would be suitable for
the case that devices have short messages
and their lengths are more or less the same.
Otherwise (i.e., the variation of message lengths
is significant between devices),
4-step approaches would be preferable.

While the main advantage of 2-step approaches
over 4-step approaches is a low 
access or transmission delay,
which makes it more suitable for URLLC,
the access delay of 2-step approaches
is strongly dependent on re-transmission strategies
due to collisions that are inevitable
because 2-step approaches are contention-based access schemes.
In particular, if a device has a data packet that has to 
be delivered with a high probability (or a low probability
of packet dropping for highly reliable transmission
requirement),
it must be re-transmitted until it is successfully transmitted
without collision.
Thus, a proper re-transmission scheme 
has to be considered to meet requirements in terms of access delay.
While there are a number of re-transmission strategies,
in order to shorten transmission or access delay in MTC,
fast retrial \cite{YJChoi06} \cite{Mutairi13}
would be promising,
in which another 
randomly selected preamble is to be immediately re-transmitted
in the next slot.

In \cite{Dai12} \cite{Zhan18} \cite{Zhang19},
performance analysis of 4-step random access approaches with
various backoff schemes 
for re-transmissions is studied in terms of access
delay. However, the delay analysis of
2-step random access approaches,
which will be simply referred to as 2-step approaches,
with fast retrial 
has not been well studied except \cite{Choi19}.
In \cite{Choi19}, 
buffered multichannel ALOHA with fast 
retrial is analyzed using
the notion of effective bandwidth and effective
capacity \cite{Chang95} \cite{Wu03},
which allows us to see the delay performance
in terms of quality-of-service (QoS) exponent
or the tail probability of queue length without analyzing
state transition probabilities in a multi-dimensional space.
As will be discussed in the paper,
since 2-step approaches with fast retrial
can be seen as 
buffered multichannel ALOHA with fast 
retrial, the results in \cite{Choi19}
can be utilized to decide key parameters
in 2-step approaches with fast retrial.
However, due to some approximations 
used in \cite{Choi19}, there exists some gap
between derived (analytic) QoS exponents and simulated ones.
Thus, in this paper, 
in order to obtain more accurate results,
the number of assumptions or approximations required for 
analysis is reduced.
Consequently, a good approximation
of the tail probability of queue length
is obtained through an analytical way, which allows us
to design 2-step approaches
with fast retrial to meet low delay requirements.

The rest of the paper is organized as follows.
In Section~\ref{S:SM}, we present the system model
for 2-step approaches 
and introduce fast retrial.
The stability and steady-state analysis are considered
in Section~\ref{S:Stab}
without any particular arrival models.
In Section~\ref{S:QoS},
based on
the notion of effective bandwidth and effective
capacity, the QoS exponent is discussed 
for independent Poisson arrivals. 
Simulation results are presented in Section~\ref{S:Sim}
and the paper is concluded in Section~\ref{S:Con}
with a few remarks.

\subsubsection*{Notation}
Matrices and vectors are denoted by upper- and lower-case
boldface letters, respectively. The superscript $\rT$ 
denotes the transpose. The support of a vector 
is denoted by ${\rm supp} (\bx)$
(which is the number of the non-zero elements of $\bx$).
$\uE[\cdot]$
and ${\rm Var}(\cdot)$
denote the statistical expectation and variance, respectively.
$\uZ = \{0,1, \ldots\}$ stands for the set of non-negative integers.

\section{System Model}	\label{S:SM}

Suppose that a system consists of a BS and 
$N$ devices, where $N \ge 1$, for MTC.
It is assumed that all devices are synchronized
(using beacon signals transmitted by the BS).
To perform random access in MTC,
devices share a pool of $L$ preambles.
The length of preambles is denoted by $M$, which is
proportional to the system bandwidth.
If a device has messages to send to the BS,
it can transmit a randomly selected preamble.
Since the number of preambles is finite,
multiple devices can choose the same preamble,
which results in preamble collision.

\begin{figure}[thb]
\begin{center}
\includegraphics[width=\figwidth]{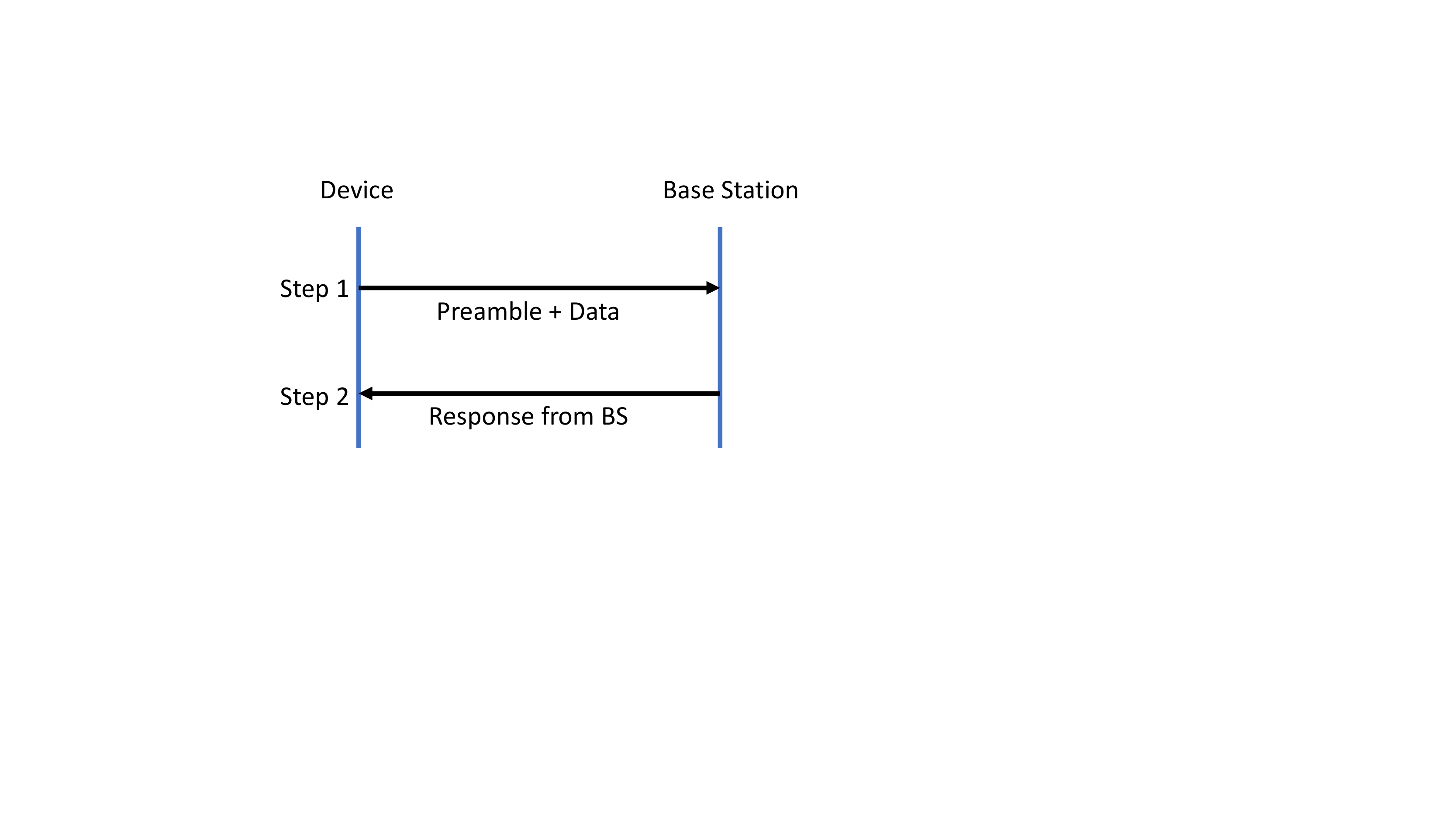}
\end{center}
\caption{An illustration of 2-step random access 
approaches in MTC.}
        \label{Fig:twostep}
\end{figure}

In this paper, we focus on 2-step approaches
for devices that need low access delay in delay-sensitive applications.
In Fig.~\ref{Fig:twostep}, an illustration of 2-step 
approaches is shown.
The first and third steps in 4-step random access
are to transmit a randomly selected preamble and a data packet, respectively,
while these two steps are combined into the first step in 2-step
random access.
In the second step, the BS is to send a feedback signal
so that devices can see whether or not collisions happen.

\begin{figure}[thb]
\begin{center}
\includegraphics[width=\figwidth]{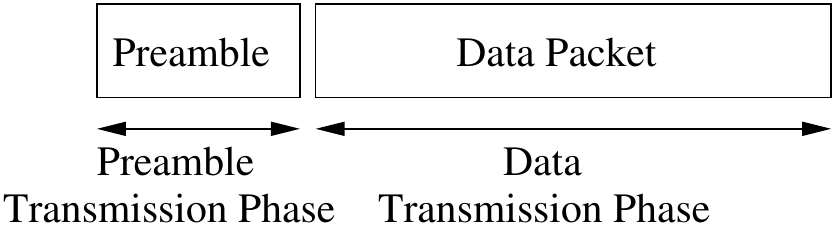}
\end{center}
\caption{Two phases
(i.e., preamble transmission and data transmission
phases) for the first step in 2-step random access.}
        \label{Fig:two_phase}
\end{figure}

The first step in 2-step random access
consists of two phases: the preamble and data transmission phases.
A randomly selected preamble is to be transmitted
in the first phase and then data packet transmission follows
in the second phase as shown in Fig~\ref{Fig:two_phase}.
In the second phase, i.e., 
the data transmission phases,
$L$ mini-slots can be considered so that an active device
that transmits preamble $l$ can transmit its data packet
in the $l$th mini-slot \cite{Choi20b}.
Alternatively, as in \cite{Abebe17}, 
$L$ spreading sequences can be used 
for data packet transmissions by multiple active devices,
where each spreading sequence is uniquely associated with a preamble.
In any case, preamble collision leads to packet collision.
For example, suppose there are $L$ mini-slots for the data transmission
phase. If multiple devices choose preamble $l$,
they will also transmit data packets in the $l$th mini-slot,
which results in packet collision. 
Thus, in this paper, preamble collision is simply referred
to as collision (as it leads to packet collision).
If a device experiences collision
(which is known by the feedback
from the BS in the second step), it can drop message.
Alternatively, it can try to re-transmit according to 
re-transmission strategies.

\begin{figure}[thb]
\begin{center}
\includegraphics[width=\figwidth]{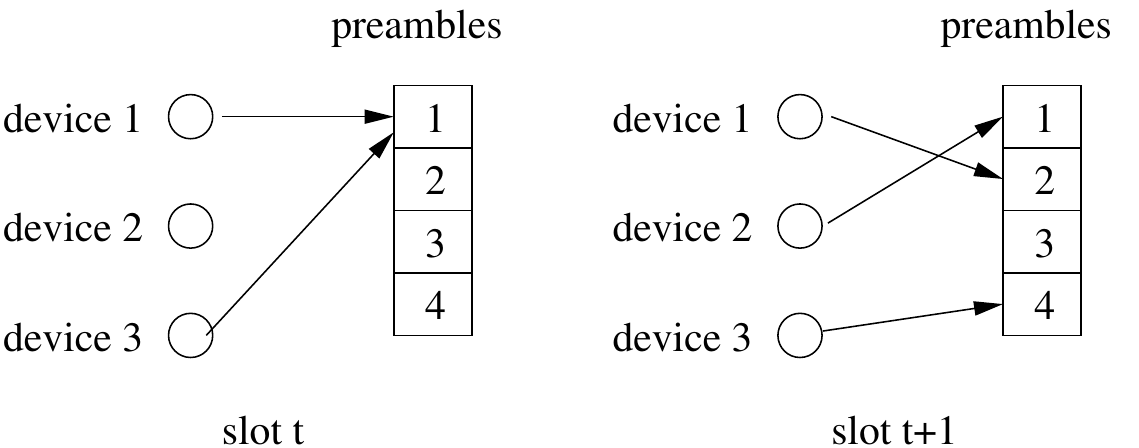}
\end{center}
\caption{An illustration
of fast retrial with 3 (type-1) devices and 4 preambles.}
        \label{Fig:frt}
\end{figure}

For re-transmissions in 2-step approaches,
in this paper, fast retrial \cite{YJChoi06} is used 
to lower access delay \cite{Mutairi13}.
With fast retrial in 2-step approaches, a device experiencing
collision immediately re-transmit with
another randomly selected preamble.
Note that as there are multiple preambles,
the resulting random access becomes multichannel ALOHA
as in \cite{YJChoi06} \cite{Mutairi13}.
Throughout the paper,
it is assumed that the two steps can be carried
out within a time slot and $t$ is used for the index
of time slots, where $t \in \{0,1, \ldots\}$.
In Fig.~\ref{Fig:frt}, we illustrate fast retrial
with $L = 4$ preambles. At slot $t$, suppose that devices 1 and 3
transmit preamble 1, which results in preamble collision.
At the next time slot, i.e., slot $t+1$,
the two devices re-transmit randomly
selected preambles (preamble 2 for device 1 and preamble 4 for
device 3), while a new active device, i.e.,
device 2, transmits preamble 1.
In this case, all the devices can successfully transmit preambles.
This shows that immediate re-transmissions by fast retrial
may not lead to successive preamble collision (unlike
single-channel ALOHA),
and shorten the access delay
(due to no back-off time).

As mentioned earlier, 
the resulting 2-step approach can be seen as
multichannel ALOHA with fast retrial.
Furthermore, for re-transmissions, each device has to have
a buffer, which results in a buffered 
(slotted) multichannel ALOHA system.

\section{Stability and Steady-State Analysis}	\label{S:Stab}

In this section, 
we study the stability of fast retrial in 2-step approaches
and steady-state analysis.
Throughout the paper, we assume that $L \ge 2$
(otherwise, fast retrial cannot be used).

\subsection{Stability}

When any re-transmission strategy (e.g., fast retrial) is employed,
each device needs to have a buffer to keep their data packets
until they can be successfully transmitted.
Denote by $q_n (t)$ the length or state
of device $n$'s queue
at time slot $t$.
Let $a_n (t)$ be the number of arrivals at 
slot $t$. For simplicity, it is assumed that 
an arrival is equal to a data block that is to send with
a preamble in the first step.
For example, if $a_n (t) = 2$, device $n$ needs at least 2 time slots
to transmit two data blocks that are arrived at time $t$,
because a device can send one data block at a time (within a time slot).
Consequently, $q_n (t) \in \{0,1, \ldots \}$
can be written as
\be
q_n (t+1) = (q_n(t) + a_n (t) - s_n(t))^+,
	\label{EQ:qq}
\ee
where $(x)^+ = \max\{0,x\}$ and
$s_n (t) \in \{0,1\}$ is the number of data blocks that can be 
possibly transmitted without collision.
That is,
\be
s_n (t)= 
\left\{
\begin{array}{ll}
1, & \mbox{if no collision happens} \cr
0, & \mbox{o.w.} \cr
\end{array}
\right.
	\label{EQ:s_n}
\ee

In general, it is expected that the length of queue,
$q_n (t)$ is finite if the system is stable.
Furthermore, for a low access delay, $q_n(t)$ has to be short,
since the access delay is proportional to the queue length.
Thus, it is important to analyze the state of queue, $q_n (t)$.
As shown in \eqref{EQ:qq}, $q_n (t)$ can be seen as a Markov chain
\cite{Norris98}.
Unfortunately, the analysis of $q_n (t)$
is not straightforward as $s_n (t)$ depends on
the other devices' queue states,
which means that the state of each queue is not independent and
the analysis of 
a $N$-dimensional Markov chain,
$\bq(t) = [q_1 (t) \ \ldots \ q_N (t)]^\rT
\in \uZ^N$ is required to take into account the interaction between
$N$ devices' queues.
In this paper, as will be explained later,
however, we do not consider a conventional approach
based on state transition probabilities
(e.g., \cite{Dai12} \cite{Zhang19}), but a simple approach
that can allow to effectively see 
the distribution of queue length individually.

Although certain steady-state analysis
has been carried out for fast retrial in 
\cite{YJChoi06} \cite{Mutairi13},
the stability was not addressed. 
In \cite{Choi18} \cite{Choi20c},
based on
Foster-Lyapunov criteria \cite{Kelly_Yudovina} \cite{HajekBook},
the stability of fast retrial has been studied.
In particular, in 
\cite{Choi20c},
the following result is obtained.

\begin{mytheorem}	\label{T:1}
Suppose that 
$a_n(t)$ is independent and identically distributed (iid) 
with a finite mean as follows:
\be
\lambda_n = \uE[a_n (t)] < \infty,
\ee
where 
$\lambda_n$ is the mean arrival rate of device $n$.
If
\be
\frac{1}{N} \sum_{n=1}^N \lambda_n 
< \left(1 - \frac{1}{L} \right)^{N-1},
	\label{EQ:T1}
\ee
then 
$\bq(t)$ 
is positive recurrent.
\end{mytheorem}

While the proof of Theorem~\ref{T:1}
can be found in \cite{Choi20c},
it is necessary to highlight that
the condition in \eqref{EQ:T1} is a sufficient condition
for stability. It can also be seen that
the right-hand side (RHS) term
in \eqref{EQ:T1} is the probability of no collision 
(or successful transmission) when all $N$ devices transmit,
which can be the worst case 
in terms of the average departure rate
as all devices compete for 
$L$ preambles.

There are a number of observations based on \eqref{EQ:T1}.
First of all, 
an asymptotic case can be considered with \eqref{EQ:T1}. Let
$\eta = \frac{N}{L}$.
With a fixed ratio $\eta$, 
as $N$ approaches $\infty$, we have 
\be
\lim_{N \to \infty} 
\left(1 - \frac{1}{L} \right)^{N-1} = e^{-\eta}.
\ee
Thus, an asymptotic version of \eqref{EQ:T1} becomes
\be
\bar \lambda <  e^{-\eta},
\ee
where $\bar \lambda = 
\lim_{N \to \infty} \frac{\sum_{n=1}^N \lambda_n}{N}$.
It can also be shown that the queues may not be stable
if $N \to \infty$ with non-zero $\lambda_n$ and a finite $L$.
That is, for a fixed $L$, 
\eqref{EQ:T1} becomes
\be
\frac{\sum_{n=1}^N \lambda_n}{N}
< e^{-\kappa_L (N-1)},
\ee
where $\kappa_L = \ln \frac{L}{L-1}$.
Thus, as $N \to \infty$, when $\bar \lambda > 0$, we see that
the 
RHS term approaches 0 with a fixed $L$.
This implies that
when the number of devices can be unbounded as in \cite{YJChoi06},
random backoff algorithms need to be used for
the case that a device fails 
to transmit after a certain number of re-transmissions with
fast retrial.
Otherwise, the average arrival rate per device
should decrease exponentially with $N$,
i.e., the decrease of $\bar \lambda$
has to be faster than $1/N$.
This shows that a large number of devices
is not desirable for 2-step approaches
with fast retrial
to ensure low access delay unless $L$ is also large.

\subsection{Steady-State Analysis}

Once the condition in \eqref{EQ:T1}
holds, a stationary distribution of $\bq(t)$ 
exists, because $\bq(t)$ is positive recurrent \cite{Norris98}.
Thus, in this
subsection, a steady-state analysis is carried out
under the assumption that 
a stationary distribution of $\bq(t)$ exists.

Let $K$ be the number of devices that send preambles 
at a slot in steady-state.
Here, the index of time slots, $t$, is omitted for convenience.
With a finite $N$, let
$\lambda = \frac{\sum_{n=1}^N \lambda_n} {N}$.
For a system in steady-state,
the average total arrival rate
is $\lambda N$, while 
the average total departure rate,
which is the average number of 
the transmitted preambles without collision, is given by
\be
G = \uE \left[ K \left(1 - \frac{1}{L} \right)^{K-1}
\right].
	\label{EQ:GE}
\ee
In steady-state, they should be the same, i.e.,
\be
\lambda N = G = \uE \left[ K \left(1 - \frac{1}{L} \right)^{K-1}
\right].
	\label{EQ:NS}
\ee

For simplicity, let
\be
\lambda_n = \lambda, \ n = 1, \ldots, N.
	\label{EQ:lln}
\ee 
Furthermore, it is assumed that 
the number of devices is larger than or equal\footnote{If $N \le L$, 
each device can have a unique preamble, which implies that
random access is not necessary.
Thus, in the paper, we only consider the case that 
$\frac{N}{L} \ge 1$.} 
to the number of preambles, i.e., $N \ge L$.
The symmetric condition in \eqref{EQ:lln}
leads to the same probability
of transmission (or access probability)
for all devices, denoted by $\alpha$.
Due to the interaction between queues,
the events that devices transmit become dependent.
However, for tractable analysis, we can consider an approximation 
that the events are independent,
which leads to the
following distribution of $K$:
\be
\Pr(K = k) = \binom{N}{k} \alpha^k (1- \alpha)^{N-k},
	\label{EQ:PK}
\ee
i.e., $K$ follows a binomial distribution.
Substituting \eqref{EQ:PK} into
\eqref{EQ:NS},
we have
\be
\lambda = S(\alpha) 
\deft  \alpha \left(1 - \frac{\alpha}{L} \right)^{N-1},
	\label{EQ:ll}
\ee
which can be used to find 
$\alpha$ from $\lambda$.

\begin{figure}[thb]
\begin{center}
\includegraphics[width=\figwidth]{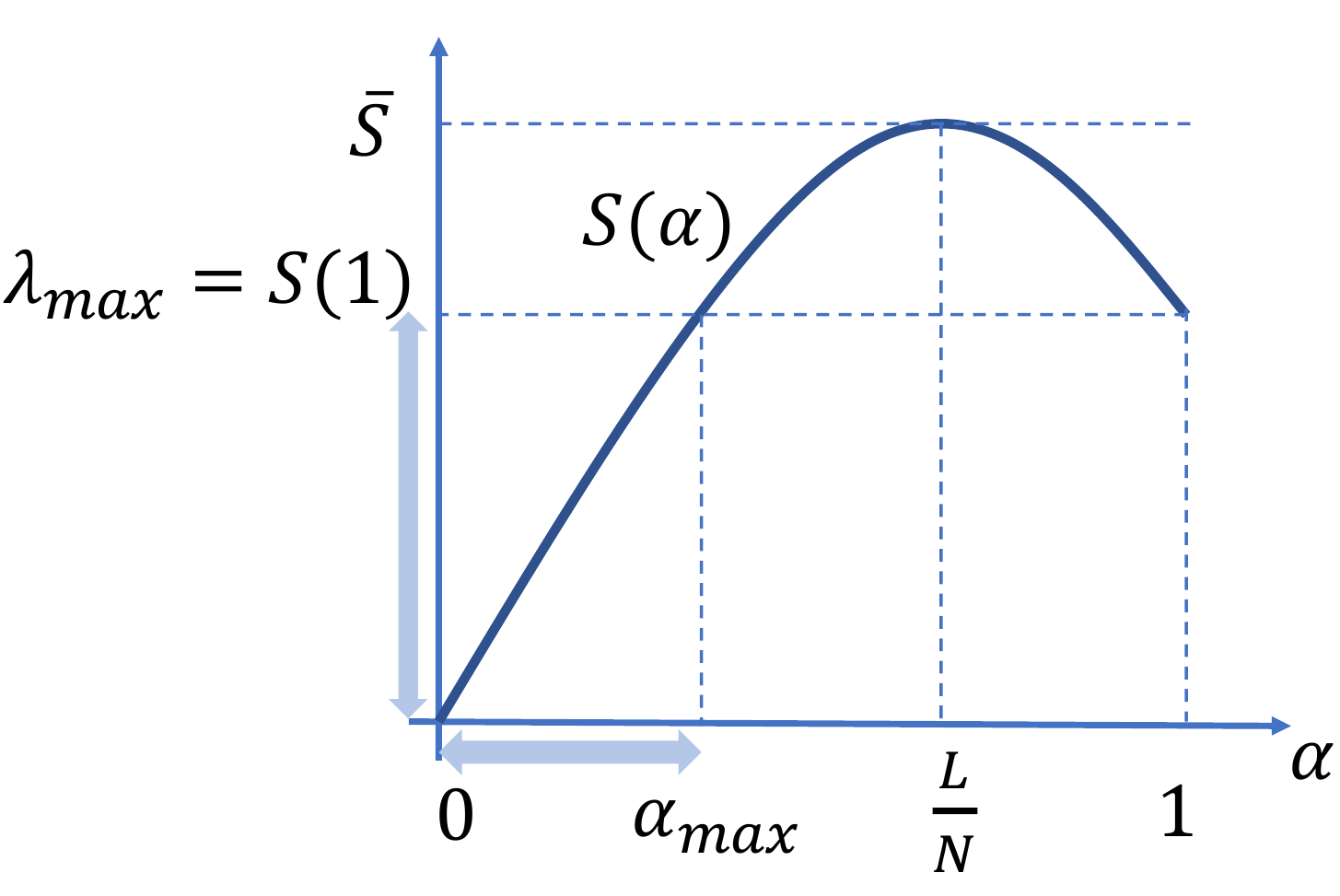}
\end{center}
\caption{An illustration of $S(\alpha)$ when $N \ge L$,
which shows that there is a unique $\alpha$ for
a given $\lambda < \lambda_{\rm max}$.}
        \label{Fig:Fig1}
\end{figure}

\begin{mylemma}	\label{L:lam_alp}
For a given $\lambda < \lambda_{\rm max}$, where 
\be
\lambda_{\rm max} = \left(1 -  \frac{1}{L} \right)^{N-1},
\ee
there is a unique $\alpha$ that satisfies \eqref{EQ:ll}.
Note that $\lambda < \lambda_{\rm max}$ implies \eqref{EQ:T1}.
\end{mylemma}
\begin{IEEEproof}
In \eqref{EQ:ll},
it can be readily shown that
$S(\alpha)$ is a unimodal function (i.e., a $\cap$-shape)
and has the maximum
when $\alpha = \frac{L}{N}$.
Thus, with $L \le N$, we have
\begin{align}
\bar S 
& = \max_{0 \le \alpha \le 1} S(\alpha) \cr
& = S \left(\frac{L}{N} \right) 
= \frac{L}{N} \left(1 - \frac{1}{N} \right)^{N-1}
\approx 
\frac{L}{N} e^{-1}.
\end{align}
As a result, if $\lambda \le \bar S$, there can be two solutions
that satisfy \eqref{EQ:ll} (because $S(\alpha)$
has a $\cap$-shape).
It can also be shown that
\be
\bar S \ge S(1) = \lambda_{\rm max}.
\ee
One of the two solutions 
that satisfy $\lambda_{\rm max} = S(\alpha)$
is obviously $\alpha = 1$ and  the other
is less than or equal to $\frac{L}{N}$,
which is denoted by $\alpha_{\rm max}$,
as shown in Fig.~\ref{Fig:Fig1}. 
In addition, for a $\lambda < \lambda_{\rm max}$,
one of two solutions is less $\alpha_{\rm max}$
and the other has to be greater than $1$. 
Since $\alpha$ is a probability, 
it has to be less than or equal to 1, i.e., $\alpha \le 1$,
which implies that there is a unique solution $\alpha$ 
that satisfies \eqref{EQ:ll} for $\lambda < \bar \lambda$.
This completes the proof.
\end{IEEEproof}

According to Lemma~\ref{L:lam_alp},
$\alpha$ can be found for a given $\lambda < \lambda_{\rm max}$.
Thus, without any difficulties,
we can define the inverse function
of $S(\alpha)$, denoted by $S^{-1} (\cdot)$, so
that $\alpha = S^{-1} (\lambda)$.
As shown in Fig.~\ref{Fig:Fig1},
clearly, $\alpha \le \alpha_{\rm max}$,
and $\alpha_{\rm max}$ is a solution 
of the following equation:
\be
S(\alpha) = 
\lambda_{\rm max} = S(1) = \left(1 - \frac{1}{L} \right)^{N-1}, 
\ \alpha \le 1,
	\label{EQ:am}
\ee
which has two solutions. One is obviously $\alpha = 1$, 
while the other is $\alpha_{\rm max} \le 1$.
In what follows,
we derive asymptotic expressions
for $\alpha_{\rm max}$  as well as 
$\alpha = S^{-1} (\lambda)$ when $\lambda < \lambda_{\rm max}$.

\begin{mylemma}	\label{L:2}
Suppose that $N \to \infty$ with a fixed ratio $\eta = \frac{N}{L}$.
Then, $\alpha_{\rm max}$ and
$\alpha = S^{-1} (\lambda)$ for $\lambda < \lambda_{\rm max}$
are given by
\begin{align}
\tilde \alpha_{\rm max} & = \lim_{N \to \infty}
\alpha_{\rm max} = - \frac{1}{\eta} \uW_0 (- \eta e^{-\eta}) \cr
\tilde \alpha & = 
\lim_{N \to \infty} \alpha  = 
\lim_{N \to \infty}
S^{-1} (\lambda)= - \frac{1}{\eta} \uW_0 (- \eta \lambda),
	\label{EQ:L2}
\end{align}
where $w = \uW_0 (x)$ is  the Lambert W function
\cite{Corless96},
which is the inverse function of $we^w$ for $w \ge -1$.
\end{mylemma}
\begin{IEEEproof}
From \eqref{EQ:am}, if $N \to \infty$, we have
$\alpha e^{ - {\alpha \eta}} =  e^{- \eta}$
or
\be
- \alpha \eta e^{ - {\alpha \eta}} =  - \eta e^{- \eta}.
	\label{EQ:aeee}
\ee
Let $w = - \alpha \eta$. Then, it becomes
$w e^w = - \eta e^{-\eta}$, which implies
that $w$ is $\uW_0 (- \eta e^{-\eta})$ or
\be
- \alpha \eta = \uW_0 (- \eta e^{-\eta}).
	\label{EQ:W0}
\ee
Clearly, \eqref{EQ:L2} is equivalent to
the first equation in \eqref{EQ:W0}.

For the second equation in \eqref{EQ:W0},
in \eqref{EQ:aeee}, the RHS term can be replaced
with $-\eta \lambda$. Then, through similar 
steps above, 
we can have the second equation in \eqref{EQ:W0},
which completes the proof.
\end{IEEEproof}

\begin{figure}[thb]
\begin{center}
\includegraphics[width=\figwidth]{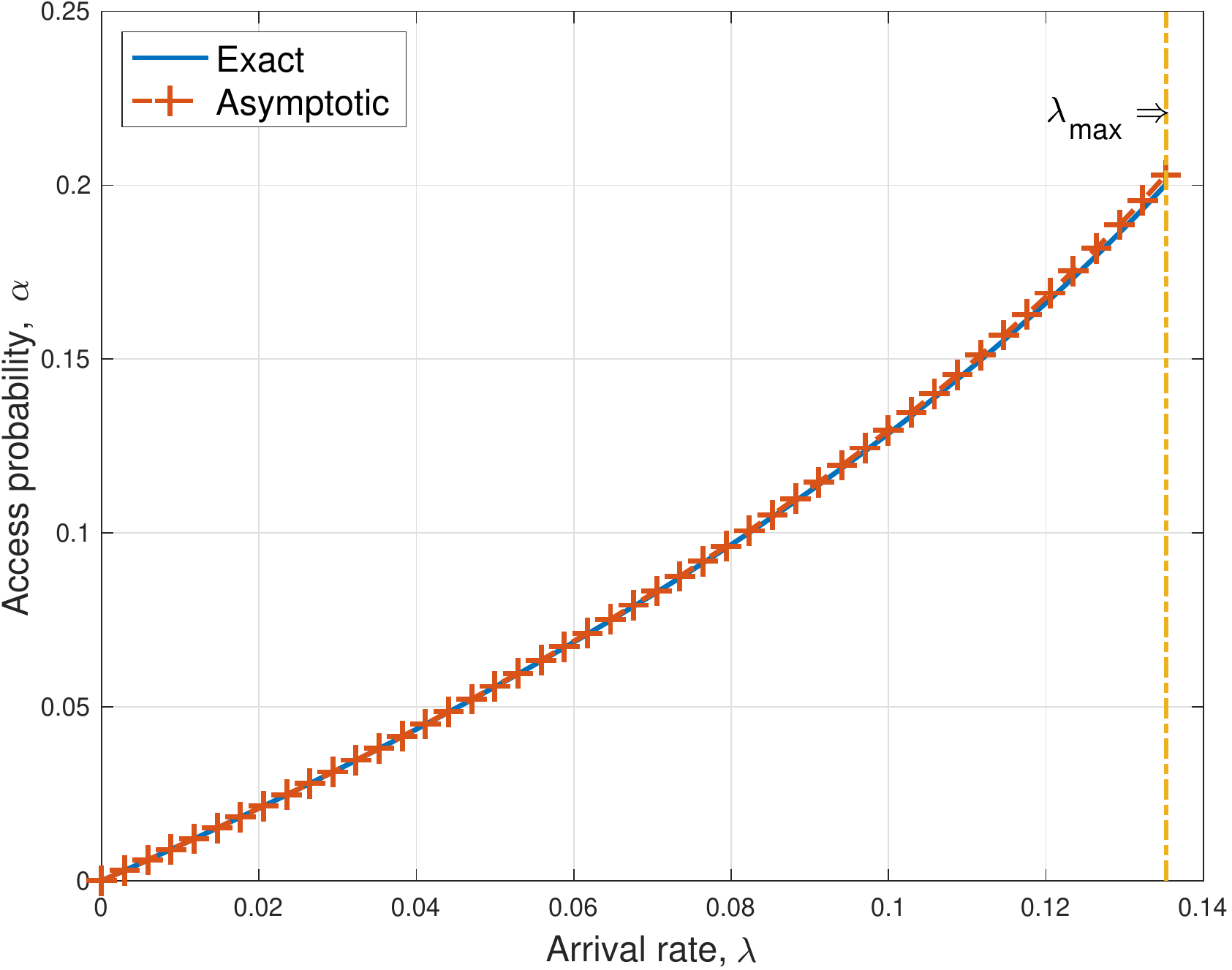}
\end{center}
\caption{The access probability, $\alpha$, as
a function of $\lambda$ when $N = 40$ and $L = 20$.}
        \label{Fig:plt_alp}
\end{figure}

In Fig.~\ref{Fig:plt_alp},
$\alpha$ is shown as a function of $\lambda$
when $N = 40$ and $L = 20$
In addition, $\tilde \alpha$, which is an asymptotic
access probability with $\eta = 2$, is shown in the figure.
It is also shown that $\tilde \alpha_{\rm max} = 0.2032$.
The difference between
$\alpha$ and $\tilde \alpha$ seems negligible 
unless $\lambda$ is close to $\lambda_{\rm max} = 
(1 - \frac{1}{L})^{N-1} = 0.1353$.

\begin{figure}[thb]
\begin{center}
\includegraphics[width=\figwidth]{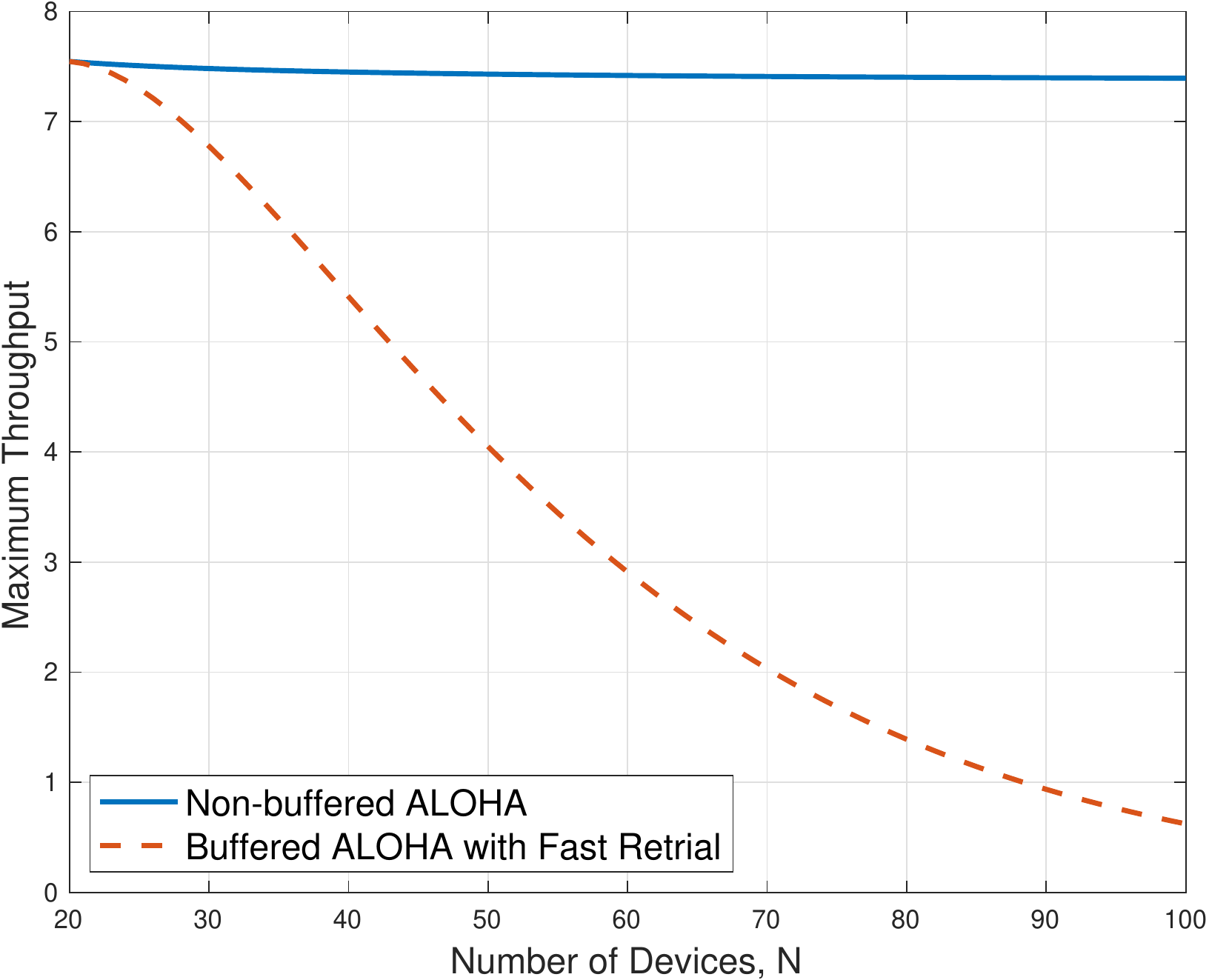}
\end{center}
\caption{Maximum throughput of conventional multichannel 
ALOHA and that of multichannel ALOHA with fast retrial
as functions of $N$ when $L = 20$.}
        \label{Fig:plt_SS}
\end{figure}

Note that since $\uW_0 (-x) \in [-1, 0]$ for $x \in [0, e^{-1}]$
and $\alpha \le 1$, it can be confirmed that
\be
\tilde
\alpha_{\rm max} \le \frac{1}{\eta} = \frac{L}{N}.
\ee
In \eqref{EQ:GE},
$G$ is usually called the throughput 
that can be maximized when
the access probability, $\alpha$, is $\frac{L}{N}$,
which is also shown in Fig.~\ref{Fig:Fig1},
where $S(\alpha) = \frac{G}{N}$ (in this sense,
$S$ can be seen as the normalized throughput per device).
In conventional multichannel ALOHA
(i.e., non-buffered multichannel ALOHA),
the system is known to be stable if $\alpha < \frac{L}{N}$
\cite{Shen03},
while multichannel ALOHA
with fast retrial 
is stable if $\alpha < \alpha_{\rm max}$.
Note that due to the presence of buffers at devices,
multichannel ALOHA with fast retrial is also 
buffered multichannel ALOHA as mentioned earlier.
In Fig.~\ref{Fig:plt_SS},
the maximum throughput of conventional 
multichannel ALOHA, which is $\max G = N \bar S$,
is shown with that of (buffered) multichannel ALOHA
with fast trial for different numbers of
devices, $N$, when $L = 20$. 
Since $\max G = N \bar S \approx L e^{-1}$, 
the maximum throughput of conventional 
multichannel ALOHA is almost independent of $N$.
On the other hand, 
the maximum throughput of multichannel ALOHA
with fast trial, $N S(1) = N \left(1 - \frac{1}{L} \right)^{N-1}$,
decreases with $N$.
Clearly, when fast retrial is employed for low
access delay, it is necessary to keep the number of devices
small.

In addition, there are few remarks as follows.
\begin{itemize}
\item In \cite{YJChoi06},
a steady-state analysis can be found 
under the assumption that the number of retrials follows
a Poisson distribution. Although this assumption simplifies
the analysis, it is not valid (as stated in \cite{YJChoi06}).
In this paper, fortunately, no assumption about the distribution of
retrials is required.

\item In \cite{Choi19}, as will be discussed in 
the next section (i.e., Section~\ref{S:QoS})
the notion of effective bandwidth is applied to
multichannel ALOHA with a number of assumptions that simplify analysis.
Among those, in this paper, 
two assumptions (in particular,
the assumptions of {\bf A1} and {\bf A2} in \cite{Choi19}) 
are not used, while one assumption (or approximation)
of independent departures 
is still used (to find the distribution of $K$
in \eqref{EQ:PK}).
Thus, the results in this paper are still approximate,
although they are reasonably close to 
simulation results as will be shown in Section~\ref{S:Sim}.
\end{itemize}

\section{QoS Exponent in Steady-State}	\label{S:QoS}

In this section, we exploit the notion of effective bandwidth
\cite{Chang95} \cite{Wu03} \cite{KellyBook} 
to find the tail distribution
of queue length in steady-state.

\subsection{Effective Bandwidth and QoS Exponent}

Consider the logarithm of moment generating functions 
(LMGFs) of
$a_n (t)$ and $s_n (t)$
that are given by
\begin{align}
\Lambda_a (\theta) & = \ln \uE[e^{\theta a_n}] \cr
\Lambda_s (\theta) & = \ln \uE[e^{\theta s_n}].
\end{align}
From \eqref{EQ:s_n},
we can also show that
\be
\Lambda_s (\theta) = \ln\left( 1 - p + p e^\theta \right), 
\ee
where $p$ is the probability of successful
transmission without collision.

In this section, we assume
that $a_n (t)$ is an independent Poisson arrival with mean $\lambda$
for all $n$. Thus, 
we have $\Lambda_a (\theta) =  \lambda (e^\theta - 1)$.
In this case, we can find a relation between 
the probability of empty queue,
i.e., $\Pr(q_n = 0)$, and the access probability
as follows.

\begin{mylemma}	\label{L:alp_q}
Under the assumption that $a_n (t)$ is Poisson,
the access probability, $\alpha$,
is given by
\be
\alpha = 1 - e^{-\lambda} \Pr(q_n = 0). 
	\label{EQ:alp_q}
\ee
\end{mylemma}
\begin{IEEEproof}
The transmission probability or access probability
is the probability that a device
has a packet to transmit. 
A device has a packet to transmit if the queue is not empty
or if there are any arrivals when the queue is empty.
Thus, it can be shown that
\begin{align}
\alpha & = \Pr(q_n \ge 1) + \Pr(a_n \ge 1 \,|\, q_n = 0)  \Pr(q_n = 0) \cr
& = \Pr(q_n \ge 1) + \Pr(a_n \ge 1)  \Pr(q_n = 0) \cr
& = (1 - \Pr(q_n = 0)) + (1- e^{-\lambda}) \Pr(q_n = 0),
\end{align}
which leads to \eqref{EQ:alp_q}.
\end{IEEEproof}

According to \cite{Chang95} \cite{Wu03},
in steady-state, the queue length
has the following tail probability:
\be
\Pr(q_n (\infty) \ge \tau) \eeq
e^{- \theta^* \tau}, \ \tau > 0,
	\label{EQ:tail}
\ee
where $\theta^* > 0$ is the solution of the 
following equation:
\be
\Lambda_a (\theta) + \Lambda_s (- \theta) = 0.
	\label{EQ:LL}
\ee
In \eqref{EQ:tail}
for two functions $f(x)$ and $g(x)$, we say that
$f(x) \eeq g(x)$ 
if $\lim_{x \to \infty} \frac{1}{x}\ln \frac{f(x)}{g(x)} = 0$,
i.e., the two functions are asymptotically 
equal to the first order in the exponent.

To find $\theta^*$, we need to know $p$.
Consider device $n$. Under the symmetric condition 
in \eqref{EQ:lln}, $p = \Pr(s_n = 1)$ for all $n$. In fact,
$s_n$ depends on the states of the other devices' queues.
For convenience, let 
$K_{-n}$ denote the number of the other devices that transmit.
Clearly, we
have $s_n = 1$ only when all the other $K_{-n}$ devices choose
different preambles.
That is,
\be
\Pr(s_n = 1\,|\, K_{-n} = k) = \left(1 - \frac{1}{L} \right)^{k}.
\ee
As mentioned earlier, if \eqref{EQ:T1} holds,
each queue has a stationary distribution.
Thus, with $\alpha$, the probability
that there are the other $k$ active devices is given by
\be
\Pr(K_{-n} = k) = \binom{N-1} {k} \alpha^k (1- \alpha)^{N-1-k}.
\ee
After some manipulations, $p$ can be found as
\begin{align}
p & = \Pr(s_n = 1) \cr
& = \sum_{k=0}^{N -1}\Pr(s_n= 1\,|\, K_{-n} = k)  \Pr(K_{-n} = k) \cr
& = \sum_{k=0}^{N -1}
\left(1 - \frac{1}{L} \right)^k 
\binom{N-1}{k} \alpha^k (1- \alpha)^{N-1-k} \cr
& = \left( 1 - \frac{\alpha}{L} \right)^{N-1}.
	\label{EQ:p_a}
\end{align}

\begin{mylemma}	\label{L:3}
Suppose that $N \to \infty$ with a fixed ratio $\eta = \frac{N}{L}$.
Then, $p$ converges to a constant that is given by
\be
\tilde p = \lim_{N \to \infty} p = e^{- \tilde \alpha \eta}
= \exp(\uW_0 (-\lambda \eta)).
	\label{EQ:L3}
\ee
\end{mylemma}
\begin{IEEEproof}
According to Lemma~\ref{L:2}, 
$\alpha$ converges to $\tilde \alpha$ as $N \to \infty$.
Thus, we can readily have \eqref{EQ:L3}.
\end{IEEEproof}

\begin{figure}[thb]
\begin{center}
\includegraphics[width=\figwidth]{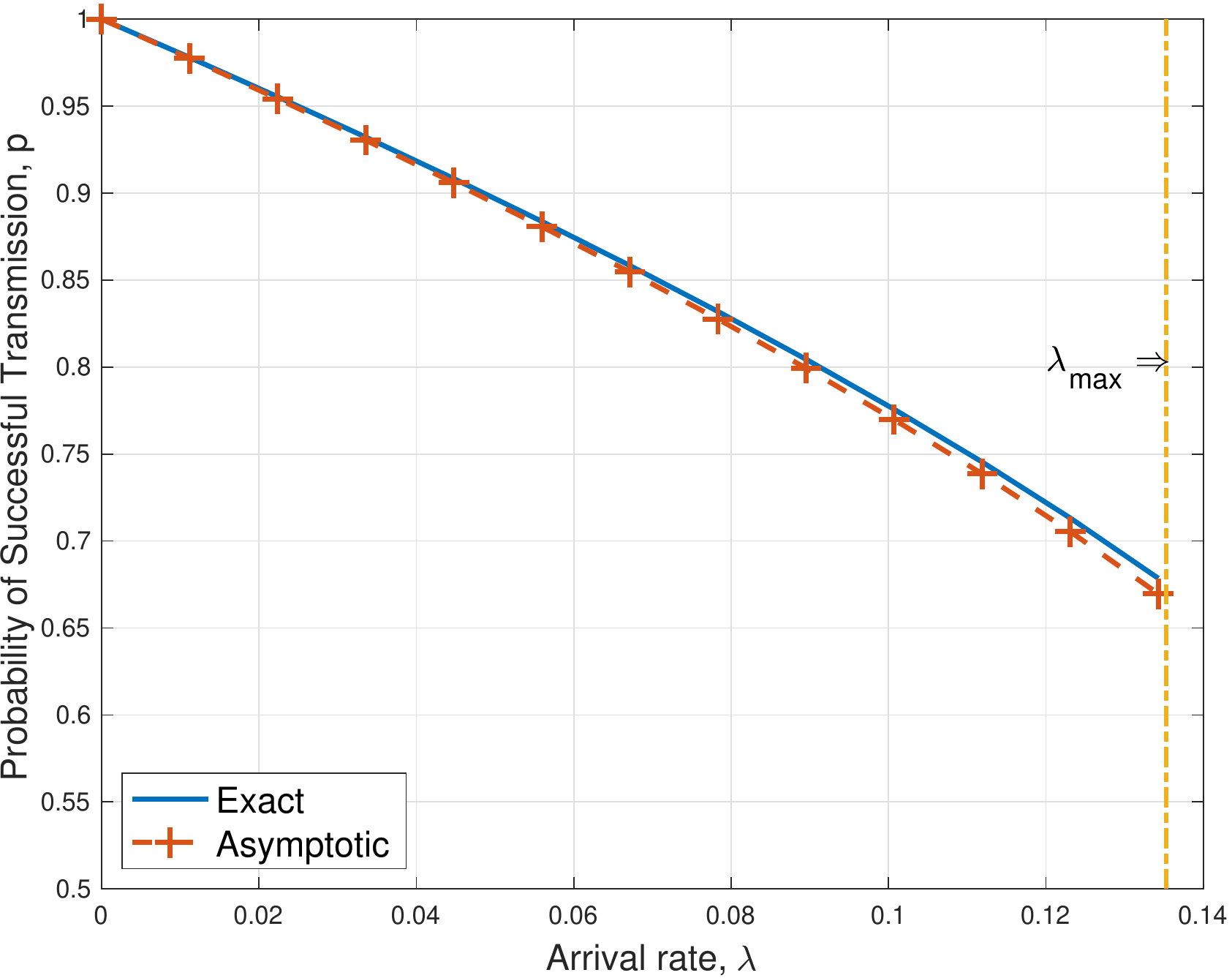}
\end{center}
\caption{The probability of successful
transmission, $p$, when $N \ge L$,
as a function $\lambda$ with its asymptotic,
$\tilde p$ when $N = 40$ and $L = 20$.}
        \label{Fig:plt_p}
\end{figure}

In Fig.~\ref{Fig:plt_p},
$p$ is shown as a function of $\lambda$
with its asymptotic, $\tilde p$, in \eqref{EQ:L3}
when $N = 40$ and $L = 20$.
The difference between $p$ and $\tilde p$
is not significant if $\lambda$ is not close to $\lambda_{\rm max}$.

Consequently, for a given $\lambda$,
from \eqref{EQ:ll}, $\alpha$ can be obtained, and then
finally, $p$ can be found from $\alpha$ using \eqref{EQ:p_a}, 
which is illustrated as follows:
$$
\lambda \to  \alpha \ (\mbox{via \eqref{EQ:ll}})
\to p \ (\mbox{via \eqref{EQ:p_a}}).
$$
Once $p$ is obtained, we can substitute it into
\eqref{EQ:LL} to find $\theta^*$, which is
the solution of the following equality from \eqref{EQ:LL}:
\be
\lambda (e^\theta - 1) + \ln
\left(
1 - p (1 - e^{-\theta})
\right) = 0.
	\label{EQ:eqty}
\ee
As a result, the asymptotic distribution 
of queue length in \eqref{EQ:tail} can be found,
which allows us to understand access delay through
\eqref{EQ:tail}.

It is also possible to set a target QoS exponent
for certain guaranteed access delay. For example,
if the probability of more than or equal to
$\tau = 3$ fast retrials
has to be less than or equal to $0.01$, we can have
\be
\theta^* = -\frac{\ln (0.01)}{\tau} = 
1.5351.
\ee
Then, for a given number of devices, $N$,
key parameters such as $\lambda$ and $L$ can be 
decided to keep $\theta^* \ge 1.5351$.
On the other hand, if $\lambda$ and $L$ are fixed, 
the number of devices, $N$, can be limited
to keep QoS in terms of the QoS exponent as part of admission control.

\subsection{Bounds on QoS Exponent and Existence of a Positive $\theta^*$}

In \cite{Choi19}, with known $p$,
bounds on $\theta^*$ are found as
\be
\theta_{\rm L} \le \theta^* \le \theta_{\rm U},
\ee
where
\begin{align}
\theta_{\rm U} = \ln \left(1- \frac{\ln (1-p)}{\lambda} \right) 
\ \mbox{and} \ 
\theta_{\rm L} = \ln \frac{p}{\lambda}.
	\label{EQ:BB}
\end{align}
In addition, it is 
shown that the solution of $\theta > 0$
that satisfies \eqref{EQ:eqty} is unique if exists.
A sufficient condition that
a positive solution, i.e.,
$\theta^*> 0$, exists is simply
$p > \lambda$,
because the lower bound,
$\theta_{\rm L}$, in
\eqref{EQ:BB} becomes positive if $p > \lambda$. 

\begin{mylemma}	\label{L:4}
Suppose that $N \to \infty$ with a fixed ratio $\eta = \frac{N}{L}
\ge 1$.
Then, if $\lambda < \lambda_{\rm max}$, $\tilde p > \lambda$.
\end{mylemma}
\begin{IEEEproof}
From \eqref{EQ:L2} and the properties of Lambert W
function \cite{Corless96}, 
it can be shown that
\be
\tilde p = \exp\left(\uW_0 (-\lambda \eta) \right) 
= \frac{\lambda \eta}{-\uW_0(-\lambda \eta)}.
\ee
Thus, $\tilde p > \lambda$ is equivalent to the following:
\be
\eta > - \uW_0(-\lambda \eta).
	\label{EQ:eW0}
\ee
Since  $\lim_{N \to \infty} \lambda_{\rm max} = e^{-\eta}$,
we have $\lambda \eta < \eta e^{-\eta}$ when $N \to \infty$,
which implies that
\be
- \uW_0(-\lambda \eta) < 
- \uW_0(-\eta e^{-\eta}) \le 
- \uW_0(-e^{-1}) = 1.
\ee
Since $\eta = \frac{N}{L} \ge 1$,
\eqref{EQ:eW0} holds, which completes the proof.
\end{IEEEproof}

According to Lemma~\ref{L:4},
we can see that $\theta^* > 0$ exists
if $\lambda < \lambda_{\rm max}$.
In other words, the stability condition
in \eqref{EQ:T1} results in a positive
QoS exponent in large systems (i.e., 
when $N \to \infty$ with a fixed ratio $\eta$).

\section{Simulation Results}	\label{S:Sim}

In this section, we present simulation results
with independent Poisson arrivals for $a_n (t)$,
i.e., $a_n (t) \sim {\rm Pois}(\lambda)$  for
all $n = 1,\ldots, N$.
In order to find the distribution of queue length,
the state of queue is obtained for
$t \in \{1,\ldots, T\}$
with $q_n(0) =0$, where $T = 4000$ 
is the number of slots in each run.
To obtain the steady-state results, we only
use the last 2000 states in each run.
Furthermore, each simulation result is an average obtained from 1000 
independent runs.
To compare with simulation results,
the tail distribution in \eqref{EQ:tail}
is used as an analytical result
with $\theta^*$ that is obtained using
$\tilde p$ for given $\lambda$ and $\eta$
using \eqref{EQ:L3},
since the asymptotic results are reasonably accurate
with not too large $N$ and $L$ 
(as shown in Figs.~\ref{Fig:plt_alp} and~\ref{Fig:plt_p}).

\begin{figure}[thb]
\begin{center}
\includegraphics[width=\figwidth]{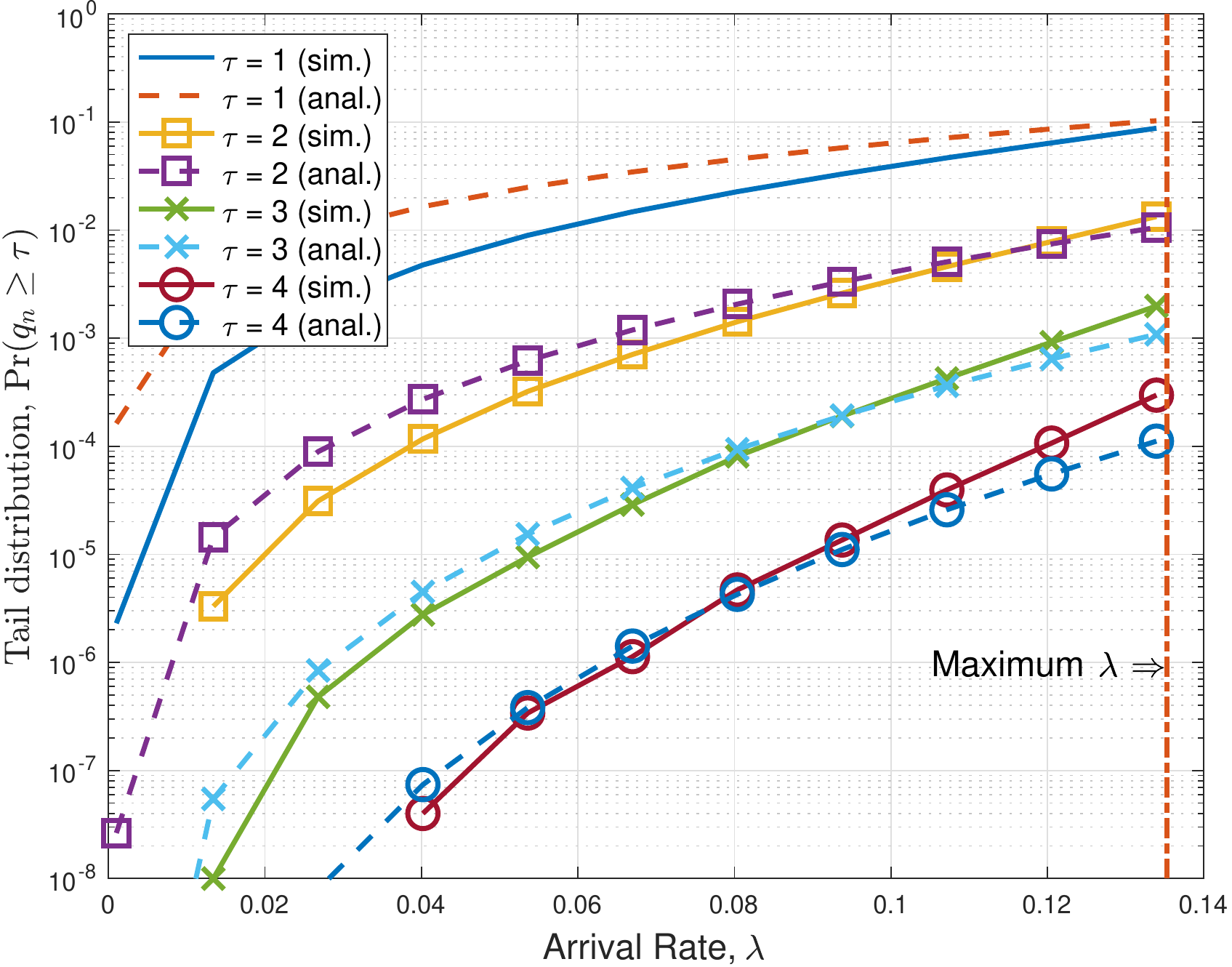} \\
(a) \\
\includegraphics[width=\figwidth]{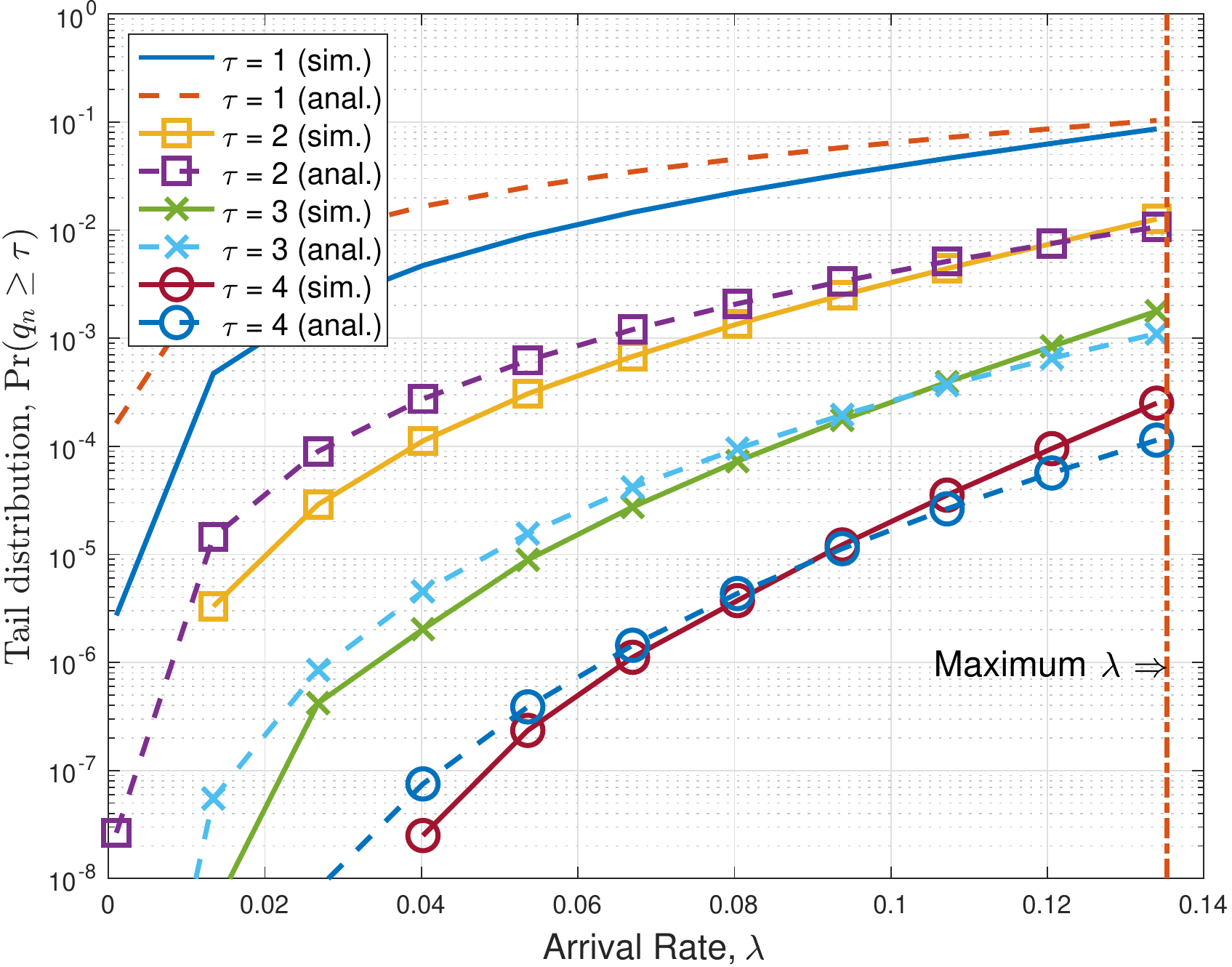} \\
(b) \\
\end{center}
\caption{Tail distribution of queue length,
$\Pr(q_n \ge \tau)$ as a function of
arrival rate, $\lambda$, with $\tau \in \{1,\ldots, 4\}$
when $\eta = 2$: (a) $N = 50$ and $L = 25$;
(b)  $N = 100$ and $L = 50$.}
        \label{Fig:plt2}
\end{figure}

Fig.~\ref{Fig:plt2}
shows the tail distribution of queue length,
$\Pr(q_n \ge \tau)$ as a function of
arrival rate, $\lambda$, with $\tau \in \{1,\ldots, 4\}$
when $\eta = 2$.
In this case, $\lambda_{\rm max}
= (1- \frac{1}{L})^{N-1} \approx e^{-\eta} = 0.1353$.
It is clear that the 
analytical results are close to simulation
results although \eqref{EQ:tail} is used for a large $\tau$.
This demonstrates that 
the analytical approach derived in this paper
can be used to design a 2-step approach with fast retrial
when access delay constraints are imposed
with certain target QoS exponents or tail probabilities.
For example, to satisfy $\Pr(q_n \ge 3) \le 10^{-4}$,
we can see that $\lambda$ has to be less
than or equal to $0.08$ 
according to Fig.~\ref{Fig:plt2}.

\begin{figure}[thb]
\begin{center}
\includegraphics[width=\figwidth]{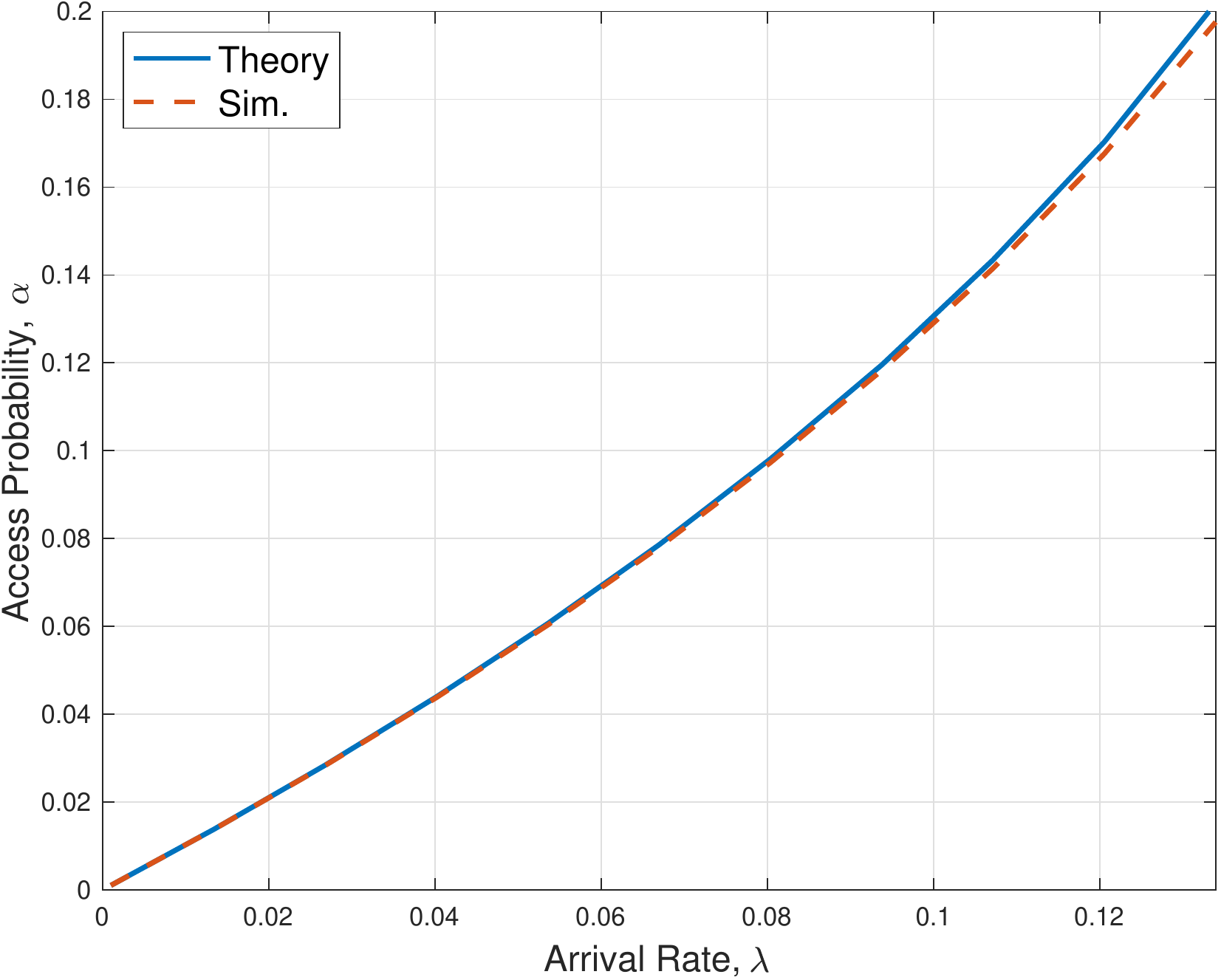} 
\end{center}
\caption{The access probability, $\alpha$, as a function
of arrival rate, $\lambda$
when $N = 50$ and $L = 25$.}
        \label{Fig:plt2_alp}
\end{figure}

In this paper, an approximation 
that the departures are independent
is used to find the distribution
of $K$ in \eqref{EQ:PK}.
For Poisson arrivals, 
we have found the relation between
$\Pr(q_n = 0)$ and $\alpha$ in \eqref{EQ:alp_q}.
From simulations, we obtain 
the probability of empty queue, i.e., $\Pr(q_n = 0)$,
and from this, a simulation
result of $\alpha$ is found using \eqref{EQ:alp_q}
for comparison with $\alpha$
obtained by solving \eqref{EQ:ll},
which is referred to as ``theory" in 
Fig.~\ref{Fig:plt2_alp}.
It is clearly shown that 
the approximation of independent departures
is reasonable to find $\alpha$
as it closely matches simulation results
(especially when $\lambda$ is small).

\begin{figure}[thb]
\begin{center}
\includegraphics[width=\figwidth]{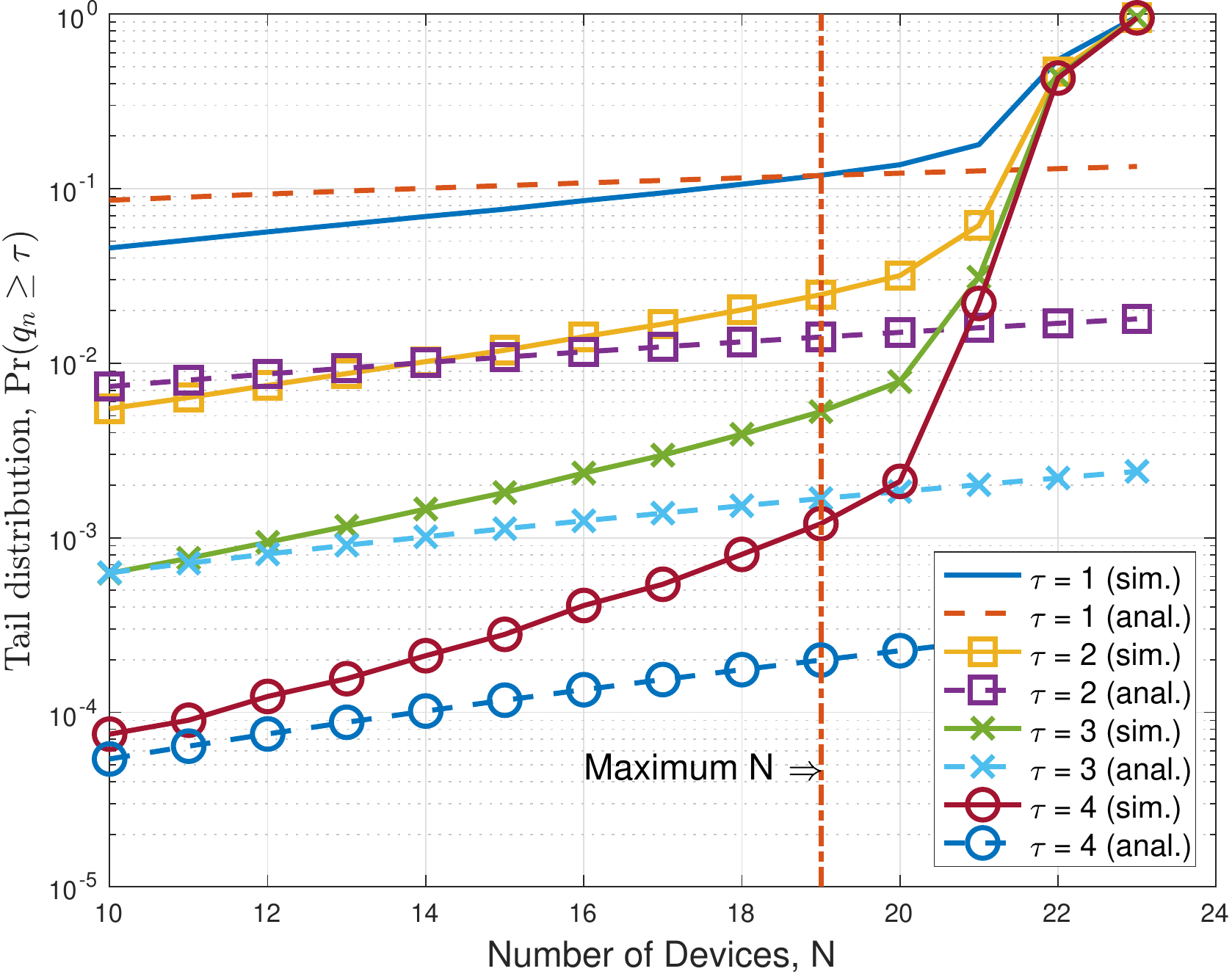} \\
(a) \\
\includegraphics[width=\figwidth]{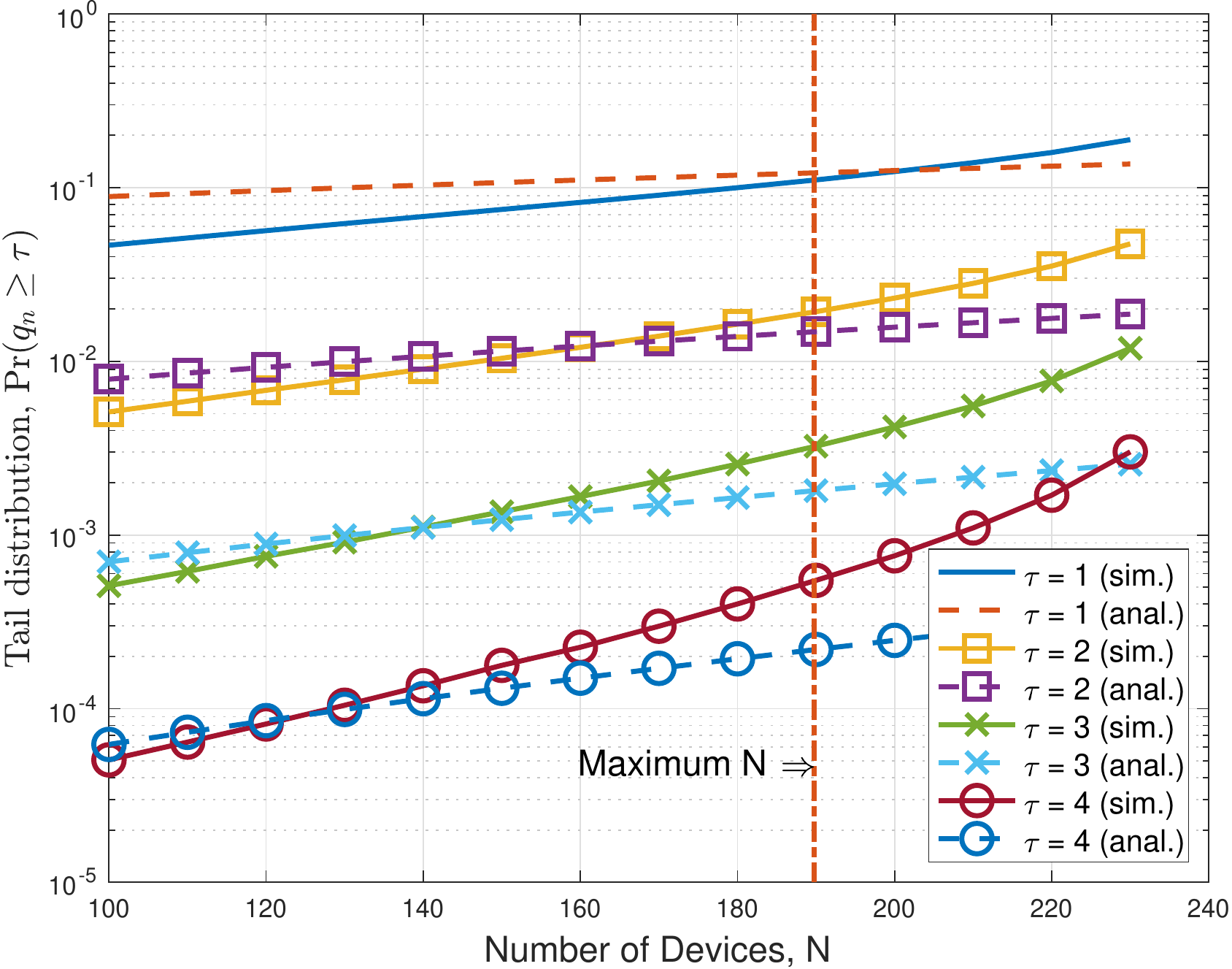} \\
(b) \\
\end{center}
\caption{Tail distribution of queue length,
$\Pr(q_n \ge \tau)$ as a function of
the number of devices, $N$, with $\tau \in \{1,\ldots, 4\}$
when $\lambda = 0.15$: (a) $L = 10$; (b) $L = 100$.}
        \label{Fig:plt1}
\end{figure}

In Fig.~\ref{Fig:plt1}, we show 
the tail distribution of queue length,
$\Pr(q_n \ge \tau)$ as a function of
the number of devices, $N$, with $\tau \in \{1,\ldots, 4\}$
when $\lambda = 0.15$. 
In general, we can see that the tail probability
increases with $N$ for a fixed $L$.
Furthermore, the analytical
results agree with simulation results
when the system is large (e.g., 
the analytical results with $L = 60$ 
(in Fig.~\ref{Fig:plt1} (b)) are closer to 
the simulation results than that with $L = 10$
(in Fig.~\ref{Fig:plt1} (a)).

It is noteworthy that for given $L$ and $\lambda$,
according to \eqref{EQ:T1},
the maximum of $N$ for a stable system,
is $N_{\rm max} = 1 + \frac{\ln \lambda}{\ln (1 - (1/L))}$.
In other words, if $N  > N_{\rm max}$,
no stability is guaranteed according to \eqref{EQ:T1}.
However, as shown in 
in Fig.~\ref{Fig:plt1}, although
$N$ is greater than $N_{\rm max}$,
the tail probability can be low.
As mentioned earlier,
since \eqref{EQ:T1} is a sufficient condition,
there can be $N > N_{\rm max}$ with stable queues.
Thus, we need to have a sufficient and necessary
condition, which will be a further issue to be studied in the future.

Furthermore, it can be observed
that with a small system (i.e., $L = 10$ in
Fig.~\ref{Fig:plt1} (a)),
the tail probability from \eqref{EQ:tail}
does not predict simulation results well when $N$
is close to $N_{\rm max}$. 
This mainly results from the 
independent departure approximation used
in \eqref{EQ:PK}, which is generally
reasonable for a large system \cite{Dai12} \cite{Choi19}.
This can be confirmed by Fig.~\ref{Fig:plt1} (b),
where $L$ is larger than that in 
Fig.~\ref{Fig:plt1} (a) by a factor of $10$,
as the tail probability from \eqref{EQ:tail}
is reasonably close to the simulation results
when $N \to N_{\rm max}$.

\begin{figure}[thb]
\begin{center}
\includegraphics[width=\figwidth]{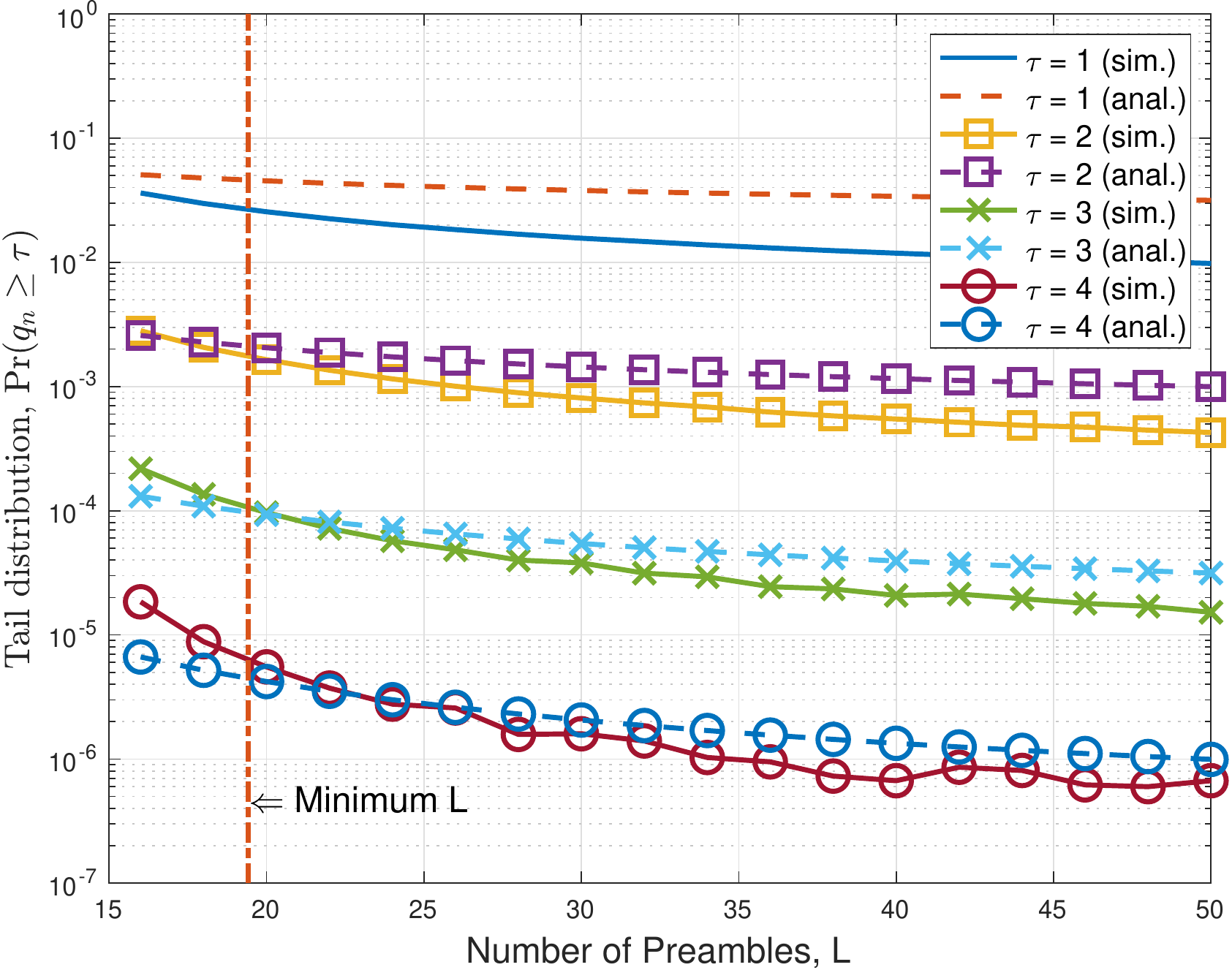} \\
(a) \\
\includegraphics[width=\figwidth]{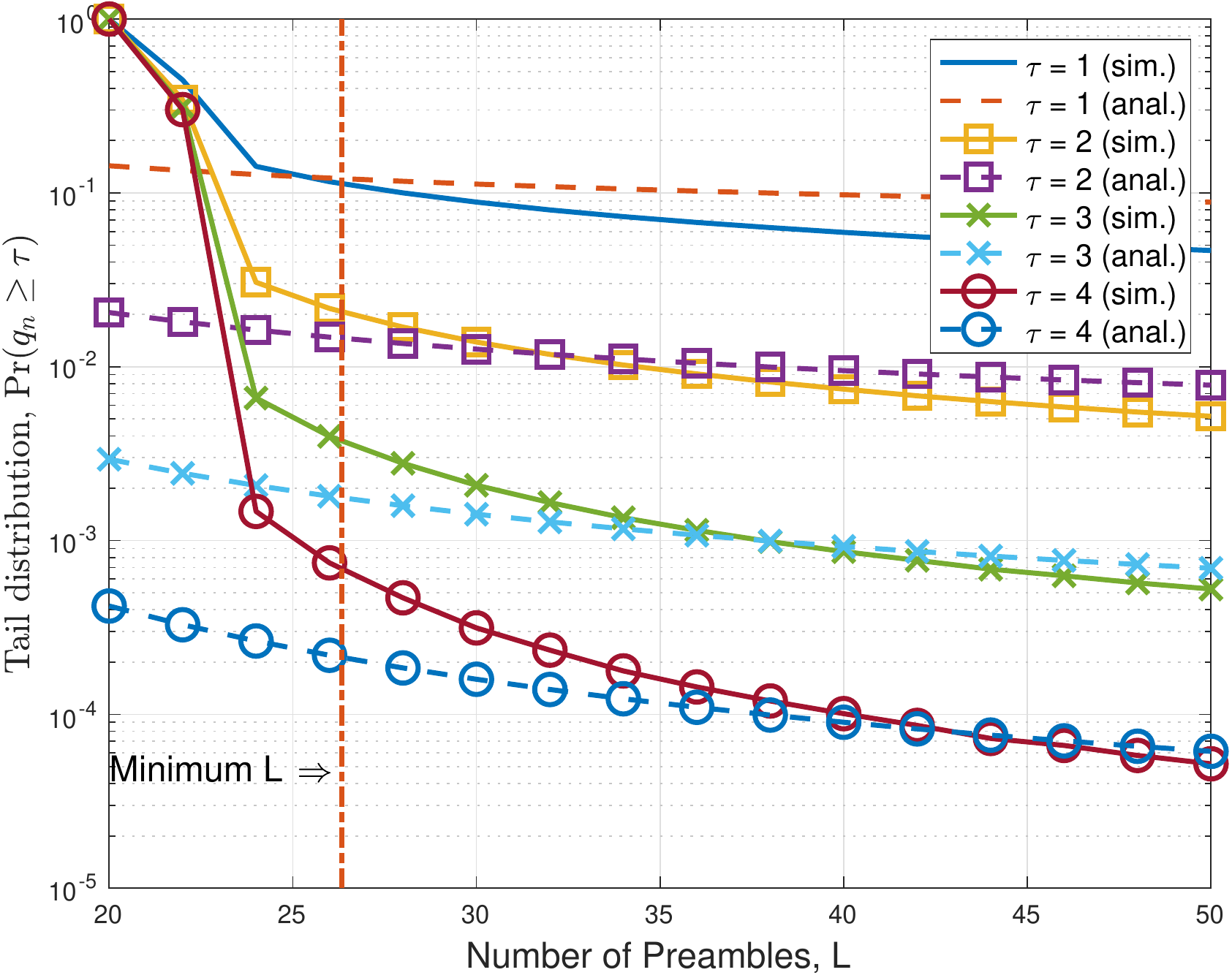} \\
(b) \\
\end{center}
\caption{Tail distribution of queue length,
$\Pr(q_n \ge \tau)$ as a function of
the number of preambles, $L$, with $\tau \in \{1,\ldots, 4\}$
when $N = 50$: (a) $\lambda = 0.075$; (b) $\lambda = 0.15$.}
        \label{Fig:plt3}
\end{figure}

Fig.~\ref{Fig:plt3} shows 
the tail distribution of queue length,
$\Pr(q_n \ge \tau)$ as a function of
the number of preambles, $L$, with $\tau \in \{1,\ldots, 4\}$
when $N = 50$ and $\lambda \in \{0.075, 0.15\}$.
It is shown that the tail probability
decreases with $L$ when $N$ is fixed.
Note that the minimum of $L$ according to \eqref{EQ:T1}
is given by $L_{\rm min} = (1 - \lambda^{\frac{1}{N-1}})^{-1}$.
However, although $L < L_{\rm min}$,
the tail probability is low according to simulation results,
which confirms again that \eqref{EQ:T1} is a sufficient condition
for stable systems.

\begin{figure}[thb]
\begin{center}
\includegraphics[width=\figwidth]{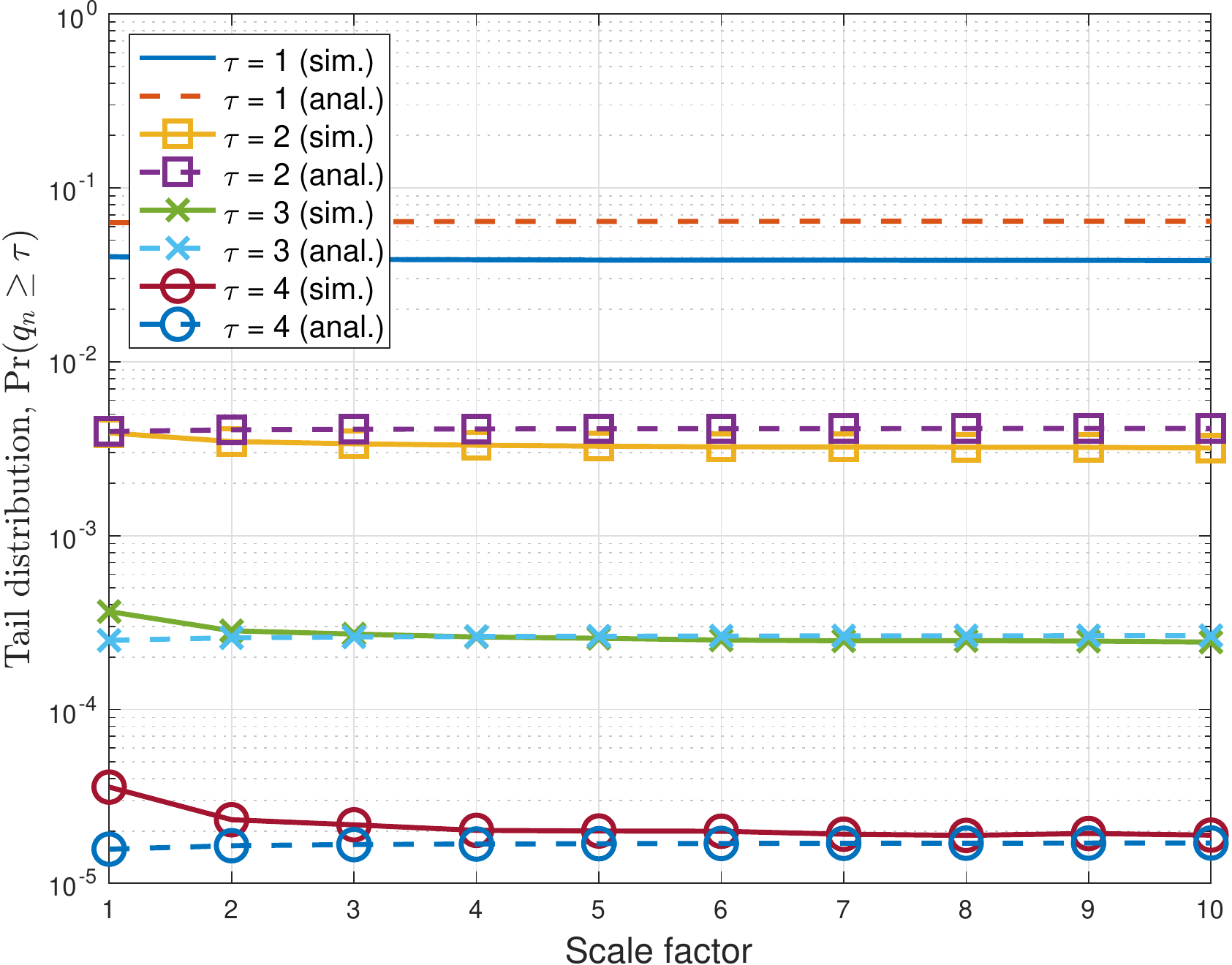} 
\end{center}
\caption{Tail distribution of queue length,
$\Pr(q_n \ge \tau)$ as a function of
the scaling factor, denoted by $s$, with $\tau \in \{1,\ldots, 4\}$
when $(N, L) = (N_0 s, L_0 s)$ and $\lambda = 0.1$.
Here, $(N_0, L_0) = (20, 10)$ for the baseline system.} 
        \label{Fig:plt4}
\end{figure}

To see the impact of the size of the system
on the performance,
we consider a scaling factor, denoted by $s$,
for a baseline system 
with $(N_0, L_0)= (20,10)$. With a fixed $\eta = 2$,
the system size increases with $s$,
i.e., $N = N_0 s$ and $L = L_0 s$.
The results are shown in 
Fig.~\ref{Fig:plt4} when $\lambda = 0.1$,
which demonstrates that
the tail probability is almost invariant with respect to
the scaling factor.
Thus, it confirms that the asymptotic analysis can be used
to understand 2-step approaches
(with large $L$ and $N$) based on
multichannel ALOHA with fast retrial.

\section{Concluding Remarks}	\label{S:Con}

In this paper, we studied 2-step random access
approaches with fast retrial in MTC 
for delay-sensitive or real-time IoT applications,
since they can have low access delay.
Under a sufficient condition for stable systems,
we performed steady-state analysis so that
the probability of successful transmission
can be analytically obtained for given parameters
such as the mean arrival rate and the numbers of devices and preambles.
Since the derived approach
in this paper required less assumptions or approximations
than \cite{Choi19},
it was able to provide a better approximation of QoS exponent,
which was confirmed by simulation results.

Based on the results in this paper,
there are a number of topics to be further 
investigated.
For example, in the case that delay-sensitive
and delay-tolerant devices co-exist,
optimal resource allocation is required.
The derived analytical approach in this paper can be used 
for optimization to 
meet target QoS exponents for delay-sensitive devices.
It is also important to improve
the derived approach in this paper for a better
approximation of QoS exponent and generalize
to the case that devices have different arrival rates.

\bibliographystyle{ieeetr}
\bibliography{mtc}

\end{document}